\documentclass[aps,prl,twocolumn,showpacs,superscriptaddress] 
{revtex4-1}
  
\usepackage{epsfig,amsmath,amssymb,bm,epsf,graphicx,psfrag}

\usepackage[all]{xy}
\usepackage{graphicx}
\usepackage{algorithm2e}
\usepackage{xcolor}
\usepackage{soul}
 
\pdfoutput=1

\newcommand{\ket}[1]{|{#1}\rangle}
\newcommand{\bra}[1]{\langle{#1}|}

\newcommand{\Tr}{\text{Tr}}

\newcommand{\hH}{\text{Herm}_1} 

\newtheorem{Theorem}{Theorem}

\newtheorem{Lemma}{Lemma}

\begin{document}

\title{Hidden Variable Model for Universal Quantum Computation with Magic States on Qubits}

\author{Michael Zurel}
\thanks{These authors contributed equally to this work}
\affiliation{Department of Physics and Astronomy, University of British Columbia, Vancouver, BC, Canada}
\affiliation{Stewart Blusson Quantum Matter Institute, University of British Columbia, Vancouver, BC, Canada}

\author{Cihan Okay}
\thanks{These authors contributed equally to this work}
\affiliation{Department of Physics and Astronomy, University of British Columbia, Vancouver, BC, Canada}
\affiliation{Stewart Blusson Quantum Matter Institute, University of British Columbia, Vancouver, BC, Canada}

\author{Robert Raussendorf}
\affiliation{Department of Physics and Astronomy, University of British Columbia, Vancouver, BC, Canada}
\affiliation{Stewart Blusson Quantum Matter Institute, University of British Columbia, Vancouver, BC, Canada}

\date{\today}

\begin{abstract}
We show that every quantum computation can be described by a probabilistic update of a probability distribution on a finite phase space. Negativity in a quasiprobability function is not required in states or operations. Our result is consistent with Gleason's Theorem and the Pusey-Barrett-Rudolph theorem.
\end{abstract}

\maketitle

It is often pointed out that the fundamental objects in quantum mechanics are  amplitudes, not probabilities \cite{QTCM,Fey}. This fact notwithstanding, here we construct a description of universal quantum computation---and hence of all quantum mechanics in finite-dimensional Hilbert spaces---in terms of a probabilistic update of a probability distribution. In this formulation, quantum algorithms look structurally akin to classical diffusion problems.

\begin{figure}[b]
\begin{center}
\includegraphics[width=8cm]{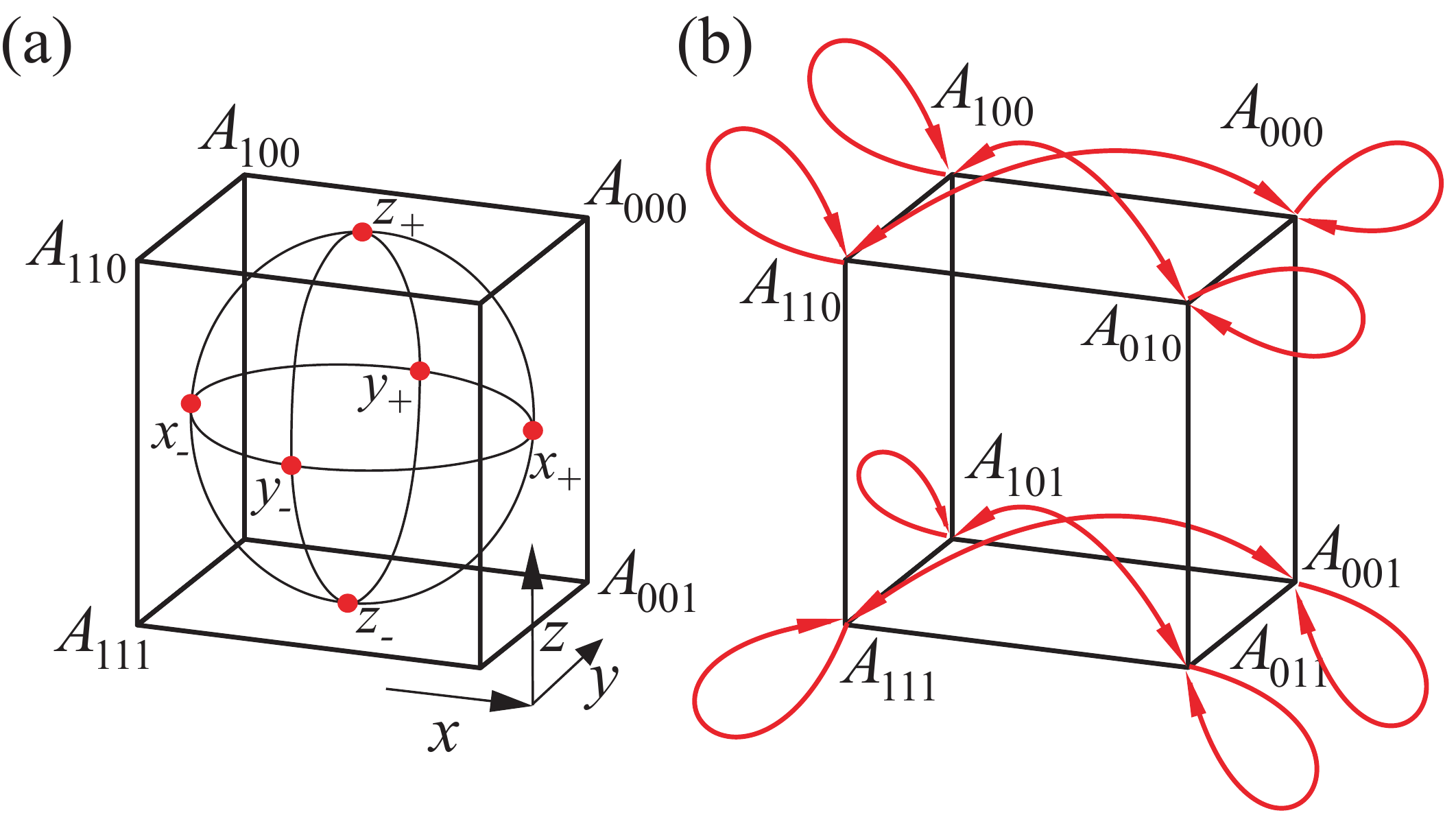}
\caption{\label{QubitHVM}
One-qubit model. (a) The state space $\Lambda_1$ is a cube with eight vertices corresponding to the phase point operators $A_\alpha=[I+(-1)^{s_x}X+(-1)^{s_y}Y+(-1)^{s_z}Z]/2$, with $\alpha=(s_x,s_y,s_z)\in\mathbb{Z}_2^3$.  The physical one-qubit states lie on or in the Bloch sphere which is contained in $\Lambda_1$ and touches the boundary of $\Lambda_1$ at six points corresponding to the six one-qubit stabilizer states. (b) Update of the phase point operators $A_\alpha$ under measurement of the Pauli observable $Z$.  Each red arrow represents a transition probability of $1/2$.}
\end{center}
\end{figure}

While this seems implausible, there exists a well-known special instance of it: quantum computation with magic states (QCM) \cite{BK} on a single qubit. Compounding two standard one-qubit Wigner functions, a hidden variable model can be constructed in which every one-qubit quantum state is positively represented \cite{Bartl2}. This representation is furthermore covariant under all one-qubit Clifford unitaries and ``positivity preserving'' under all one-qubit Pauli measurements. The update under such operations preserves the probabilistic character of the model, and hence QCM on one qubit can be classically simulated by a probabilistic update of a probability function on eight elements (see Fig.~\ref{QubitHVM} for illustration).

The prevailing view on the one-qubit example is that it is an exception and that for multiple qubits negativity will inevitably creep into any quasiprobability function of any computationally useful quantum state, rendering classical simulations inefficient \cite{Pasha}. This hypothesis is informed by the study of Wigner functions in finite-dimensional state spaces, which establishes Wigner function negativity as a necessary computational resource, i.e., there can be no quantum speedup without negativity~\cite{Gross,Galv1,Galv2,NegWi,ReWi,Delf2,QuWi,RoM,QuWi19,Heinrich,Kara,Zhu,Love,Brot,Brot2}. A quantum optics notion of quantumness---negativity of Wigner functions~\cite{Hudson1974,KenfackZyczkowski2004}---and a computational notion---hardness of classical simulation---thus align.

The viewpoint just summarized requires correction. As we show in this Letter, the one-qubit case is not an exception; rather it is an example illustrating the general case. {\em{Every}} quantum state on any number of qubits can be represented by a probability function, and the update of this probability function under Pauli measurement is also probabilistic. This is the content of Theorem~\ref{MT} below. We emphasize that the states and operations are {\em{both}} represented positively, not just one or the other.

In Theorem~\ref{L3}, we apply this to quantum computation with magic states, showing that universal quantum computation can be classically simulated by the probabilistic update of a probability distribution.

This looks all very classical, and therein lies a puzzle. In fact, our Theorem 2 is running up against a number of no-go theorems: Theorem~2 in \cite{Ferrie} and the Pusey-Barrett-Rudolph (PBR) theorem \cite{PBR} say that probability representations for quantum mechanics do not exist, and \cite{NegWi,ReWi,Delf2,QuWi,QuWi19} show that negativity in certain Wigner functions is a precondition for speedup in quantum computation. Further, does not Gleason's theorem prove that the proper representation of state in quantum mechanics is density matrices rather than probability distributions?

As we explain in the discussion part of this Letter, there is no contradiction with those works. Rather, the above-quoted theorems make stronger assumptions than we do and establish no-go theorems because of that.  However, for describing universal quantum computation---hence all quantum mechanics in finite-dimensional Hilbert spaces---our weaker assumptions suffice.

The remainder of this Letter is organized as follows. First, we define our setting and state our main results, Theorems~\ref{MT} and \ref{L3}. Then we prove them, and thereafter discuss the above questions. \smallskip

{\em{Setting and Results.}}---We focus on systems of $n$ qubits for any $n\in \mathbb{N}$ (the statement below applies to qudits in an analogous manner) and consider arbitrary quantum states evolving under sequences of Pauli measurements. The measurements need not commute, and the sequences may be arbitrarily long. This setting comprises universal quantum computation with magic states. 

Below we devise a probability representation for this setting. The representation lives on a finite generalized phase space, and its purpose is to correctly reproduce the joint measurements statistics for all quantum states and all sequences of Pauli measurements.

We denote the $n$-qubit Pauli operators by $T_a = i^{\phi(a)}X(a_X)Z(a_Z)$, $\forall a=(a_X,a_Z) \in \mathbb{Z}_2^n \times \mathbb{Z}_2^n=:E_n$, with $X(a_X):= \mathop{\otimes}\limits_{k=1}^n (X_k)^{a_X[k]}$, etc. The phases $\phi: E_n\longrightarrow \mathbb{Z}_4$ are free to choose, subject to the constraint that all $T_a$ are Hermitian. The projectors onto the eigenspaces of Pauli observables are $\Pi_{a,s} := [I+(-1)^sT_a]/2$.

The state space $\Lambda_n$ of our probabilistic model is defined as follows. We denote by $\hH(2^n)$ the set of Hermitian operators on $n$-qubit Hilbert space $H=\mathbb{C}^{2^n}$ with the property that $\text{Tr}(X)=1$ for all $X\in \hH(2^n)$, and by ${\cal{S}}_n$ the set of all $n$-qubit pure stabilizer states~\cite{Goma,Goma2,NieCha}.

Then, we define the polytope $\Lambda_n$ as
\begin{equation}\label{CS2}
\Lambda_n:=\{X \in \hH(2^n)|\, \text{Tr}(|\sigma \rangle \langle \sigma | X) \geq 0,\; \forall \, |\sigma \rangle \in {\cal{S}}_n \}
\end{equation}
(also see \cite{Heim}).  The elements $X\in \Lambda_n$ are the ``states'' (though not necessarily proper quantum states) that behave ``well'' under all sequences of Pauli measurements; namely, the probabilities for the outcome sequences are all non-negative and add up to unity. 

$\Lambda_n$ is defined as the intersection of a finite number of halfspaces. Furthermore, it is bounded [see Supplemental Material (SM), Section VI]. Therefore, by the Minkowski-Weyl theorem \cite{Ziegler1995,LP}, $\Lambda_n$ can equivalently be described as the convex hull of finitely many extreme points (vertices).  Denote by ${\cal{A}}_n$ the set of vertices of $\Lambda_n$, and the vertices by $A_\alpha \in {\cal{A}}_n$. These are our generalized phase point operators, and the corresponding index set $\{\alpha\}=:\mathcal{V}_n$ is the generalized phase space. 

We now have the following result.
\begin{Theorem}\label{MT}
For all numbers of qubits $n\in \mathbb{N}$, (i) each $n$-qubit quantum state $\rho$ can be represented by a probability function $p_\rho: \mathcal{V}_n \longrightarrow \mathbb{R}_{\ge0}$,
\begin{equation}\label{Def_p}
\rho = \sum_{\alpha\in\mathcal{V}_n} p_\rho(\alpha)\, A_\alpha.
\end{equation}

(ii) For the state update under Pauli measurements it holds that
\begin{equation}\label{PAP}
\Pi_{a,s} A_\alpha \Pi_{a,s} = \sum_{\beta \in \mathcal{V}_n} q_{\alpha,a}(\beta,s)\, A_\beta.
\end{equation}
For all $a \in E_n$, $\alpha\in \mathcal{V}_n$, the  $q_{\alpha,a}: \mathcal{V}_n\times \mathbb{Z}_2\longrightarrow \mathbb{R}_{\ge0}$ are probability functions, 

(iii) Denote by $P_{\rho,a}(s)$ the probability of obtaining outcome $s$ for a measurement of $T_a$ on the state $\rho$. Then,
the Born rule $P_{\rho,a}(s)=\text{{\em{Tr}}}(\Pi_{a,s}\rho)$ takes the form
\begin{equation}\label{BR}
	\text{{\em{Tr}}}(\Pi_{a,s}\rho) = \sum_{\alpha \in {\cal{V}}_n}p_\rho(\alpha)Q_a(s|\alpha),
\end{equation}
where $Q_a(s|\alpha)$ is given by 
\begin{equation}\label{Qq_rel}
Q_a(s|\alpha) :=\sum_{\beta \in \mathcal{V}_n} q_{\alpha,a}(\beta,s).
\end{equation}
Hence $0\leq Q_a(s|\alpha) \leq 1$, for all $a,s,\alpha$.
\end{Theorem}
The above theorem describes a hidden variable model (HVM) \cite{EPR,Bell,KS,Bohm,BB}. For any fixed number of qubits, any quantum state can be described by a probability function with finitely many elements. This property distinguishes it from the HVM of Beltrametti and Bugajski \cite{BB}, which also applies to all quantum states but requires an infinite state space. A further distinguishing property is the probabilistic state update under a dynamical process, Pauli measurement.

Theorem~\ref{MT} is illustrated in Fig.~\ref{QubitHVM} through the example of a single qubit and in the SM, Section V, for two qubits.

Because of its capability to describe Pauli measurements, the above HVM has bearing on a model of universal quantum computation, namely quantum computation with magic states (QCM) \cite{BK}. QCM is closely related to the circuit model (see the SM, Sec. IV, for background). The difference is that in QCM the set of operations is restricted to a sequence of Clifford unitaries interspersed with Pauli measurements. These operations are applied to an initial ``magic'' state. The only property of QCM relevant for the present discussion is its quantum computational universality \cite{BK,NieCha,Yao}.

To apply the above probabilistic representation to QCM, we need to consider all its operational primitives---the Pauli measurements, the Clifford unitaries, and the magic states. Magic states (like all other quantum states) and Pauli measurements are positively represented by our HVM, cf. Theorem~\ref{MT}. This leaves the Clifford gates. The easiest way of dealing with them is to observe that they are redundant, i.e., no computational power is lost if we consider sequences of Pauli measurements only. The reason is that the Clifford unitaries may be propagated past all measurements, thereby conjugating the Pauli measurements into (other) Pauli measurements. After forward propagation, the unitaries can be dropped since they do not affect the statistics of the (now earlier) measurements (see e.g. \cite{ReWi,QuWi}).

With the Pauli measurements  as the only essential dynamical element, QCM matches the setting described in Theorem~\ref{MT}. This leads to the following result.

\begin{algorithm}[h]
\parbox{8cm}{
\begin{enumerate}
\item{Sample from the probability distribution $p_\rho$. Obtain a phase space point $\alpha_0 \in \mathcal{V}_n$.}
\item{For all Pauli measurements $T_{a_t}$, $a_t \in E_n$, from $t=1$ to $t=t_\text{max}$, sample from $q_{\alpha,{a=a_t}}$ to obtain the new phase space point $\beta \in \mathcal{V}_n$ and measurement outcome $s$. Output $s$ as the outcome for the measurement of $T_{a_t}$, update the phase space point $\alpha_{t-1} \rightarrow \alpha_t=\beta$, and increment $t\rightarrow t+1$.}
\end{enumerate}}
\caption{\label{Sampling} Classical procedure to simulate a single run of a given QCM.}
\end{algorithm}

\begin{Theorem}\label{L3}
For any $n\in \mathbb{N}$ and all $n$-qubit quantum states $\rho$ the classical algorithm of Algorithm~\ref{Sampling} for sampling the outcomes of any sequence of Pauli measurements on $\rho$ agrees with the predictions of quantum mechanics.
\end{Theorem}
Thus, the HVM of Theorem~\ref{MT} describes all of universal quantum computation and hence arbitrarily closely approximates all quantum mechanical dynamics in finite-dimensional Hilbert spaces.

Theorem~\ref{L3} does not imply that the classical simulation algorithm of Table~\ref{Sampling} is efficient.  Intuition derived from previous classical simulation algorithms for quantum computation~\cite{NegWi,QuWi19,Vidal2003,Shep} suggests that it is inefficient.  However, at present we can neither prove the efficiency nor the inefficiency of this algorithm.

\smallskip

{\em{Proofs.}}---We now turn to the proofs of Theorems~\ref{MT} and \ref{L3}. The proof of Theorem~\ref{MT} requires a lemma.
\begin{Lemma}\label{L1}
The set $\Lambda_n$ has the following properties.
\begin{enumerate}
\item[(1)]{$\Lambda_n$ contains all $n$-qubit quantum states; i.e., for all $n$-qubit density operators $\rho$ it holds that $\rho \in \Lambda_n$.}
\item[(2)]{$\Lambda_n$ is closed under Pauli measurement, i.e., for all $\Pi_{a,s}$ it holds that
$$
X\in \Lambda_n\; \wedge\; \text{{\em{Tr}}}(\Pi_{a,s}X)>0 \Longrightarrow \frac{\Pi_{a,s}X\Pi_{a,s}}{\text{{\em{Tr}}}(\Pi_{a,s}X)} \in \Lambda_n.
$$} 
\end{enumerate}
\end{Lemma}
{\em{Proof of Lemma~\ref{L1}.}} All quantum states $\rho$ satisfy the conditions $\text{Tr}\big( |\sigma\rangle \langle \sigma| \rho \big)\geq 0$, for all $n$-qubit stabilizer states $|\sigma\rangle$ (as well as all other pure states), and $\text{Tr}(\rho)=1$; hence all quantum states $\rho$ are in $\Lambda_n$.

Regarding Property 2, we observe that  for all stabilizer states $|\sigma\rangle \in {\cal{S}}_n$ and all Pauli observables $T_a$ it holds that
\begin{equation}\label{P_prop}
\Pi_{a,s} |\sigma\rangle \langle \sigma| \Pi_{a,s} = c\, |\sigma'\rangle \langle \sigma'|,\; \text{where }|\sigma'\rangle \in {\cal{S}}_n,\; c\geq 0.
\end{equation}
Namely, $c=1$ if $(-1)^sT_a |\sigma\rangle= |\sigma\rangle$, $c=0$ if $(-1)^sT_a |\sigma\rangle= -|\sigma\rangle$, and $c=1/2$ otherwise \cite{NieCha}. Combining Eq.~(\ref{P_prop}) and the definition of $\Lambda_n$, Eq.~(\ref{CS2}), $\Tr(\ket{\sigma}\bra{\sigma}\Pi_{a,s}X\Pi_{a,s})=\Tr[(\Pi_{a,s}\ket{\sigma}\bra{\sigma}\Pi_{a,s})X]=c\,\Tr(\ket{\sigma'}\bra{\sigma'}X)\ge0$.
Therefore, whenever $\Tr(\Pi_{a,s}X)>0$, the post-measurement state $X'_{a,s}:=\Pi_{a,s}X\Pi_{a,s}/\text{Tr}(\Pi_{a,s}X)$ also has the property that
$$
\Tr(|\sigma\rangle \langle \sigma |X'_{a,s})\geq 0, \;\; \forall a\in E_n,\,\forall s\in\mathbb{Z}_2,\,\forall |\sigma\rangle \in {\cal{S}}_n.
$$
Furthermore, $\Tr(X'_{a,s})=1$. Therefore, $X'_{a,s} \in \Lambda_n$. $\Box$\smallskip

{\em{Proof of Theorem~\ref{MT}.}} With Property 1 in Lemma~\ref{L1}, any $n$-qubit quantum state $\rho$ is in $\Lambda_n$. Hence it can be expressed as a convex combination of the vertices $A_\alpha$, as in Eq.~(\ref{Def_p}). Taking the trace of Eq.~(\ref{Def_p}) yields $\sum_\alpha p_\rho(\alpha)=1$, i.e., $p_\rho$ is a probability function. This proves the first statement of Theorem~\ref{MT}.

With Property 2 of Lemma~\ref{L1}, for all phase point operators $A_\alpha$ and all projectors $\Pi_{a,s}$ with $\Tr(\Pi_{a,s} A_\alpha)>0$ it holds that $\Pi_{a,s}A_\alpha \Pi_{a,s}/ \text{Tr}(\Pi_{a,s}A_\alpha) \in \Lambda_n$. Therefore,
$$
\Pi_{a,s} A_\alpha \Pi_{a,s} =  \sum_{\beta\in\mathcal{V}_n} q_{\alpha,a}(\beta,s) A_\beta,
$$
with $q_{\alpha,a}(\beta,s)\geq 0$ for all $\beta \in \mathcal{V}_n$, and $s\in \mathbb{Z}_2$. Now fixing $\alpha, a$ and adding the corresponding equations for $s=0$ and $s=1$, and then taking the trace, we find
\begin{equation}\label{2sums}
\sum_{s\in \mathbb{Z}_2}\sum_{\beta\in \mathcal{V}_n} q_{\alpha,a}(\beta,s) = 1.
\end{equation}
Hence, $q_{\alpha,a}:{\cal{A}}_n\times \mathbb{Z}_2 \longrightarrow \mathbb{R}_{\ge0}$ is a probability distribution for all $\alpha \in \mathcal{V}_n$, $a\in E_n$. This demonstrates Eq.~(\ref{PAP}).

Regarding $Q_a(s|\alpha)$ as defined in Eq.~(\ref{Qq_rel}), since the $q_{\alpha,a}(\beta,s)$ are all positive, it holds that $Q_a(s|\alpha)\geq 0$ for all $a,s,\alpha$. Furthermore, with Eq.~(\ref{2sums}) it follows that $Q_a(0|\alpha)+Q_a(1|\alpha)=1$ for all $a,\alpha$, and therefore
$$
0 \leq Q_a(s|\alpha) \leq 1, \;\; \forall a,s,\alpha.
$$
Combining Eq. (\ref{Def_p}) and the already established Eq.~(\ref{PAP}),
\begin{align*}
	\Tr(\Pi_{a,s}\rho)&=\sum\limits_{\alpha\in\mathcal{V}_n}p_\rho(\alpha)\Tr(\Pi_{a,s}A_\alpha\Pi_{a,s})\\
	&=\sum\limits_{\alpha\in \mathcal{V}_n}p_\rho(\alpha)\sum\limits_{\beta\in\mathcal{V}_n}q_{\alpha,a}(\beta,s)\\
	&=\sum\limits_{\alpha\in\mathcal{V}_n}p_\rho(\alpha)Q_a(s|\alpha).
\end{align*}
This proves the formulation Eq.~(\ref{BR}) of the Born rule. $\Box$\smallskip

{\em{Proof of Theorem~\ref{L3}}.} Consider a Pauli measurement $T_a$ on input state $\rho$. Using the classical simulation algorithm, the conditional probability of obtaining outcome $s$ given the state $\alpha\in \mathcal{V}_n$ is given by Eq.~(\ref{Qq_rel}).
Therefore, the probability of obtaining outcome $s$ given a measurement of $T_a$ on state $\rho$ as predicted by the classical simulation algorithm is
\begin{equation}\label{eq:classicalProbability}
	P_{\rho,a}^{(Sim)}(s)=\sum\limits_{\alpha\in \mathcal{V}_n}p_\rho(\alpha)Q_a(s|\alpha).
\end{equation}
The outcome probability predicted by the Born rule, $P_{\rho,a}^{(QM)}$, is given by Eq.~(\ref{BR}).

Comparing Eq.~(\ref{eq:classicalProbability}) and Eq.~(\ref{BR}), we see that the classical simulation algorithm reproduces the outcome probabilities predicted by the Born rule for a single Pauli measurement.

Now we turn to the post-measurement state $\rho'$.  Quantum mechanics predicts it to be
$$
\rho'^{(QM)}=\frac{\Pi_{a,s}\rho \Pi_{a,s}}{\text{Tr}(\Pi_{a,s}\rho)}.
$$
Here the numerator is
\begin{align*}
	\Pi_{a,s}\rho\Pi_{a,s}&=\sum\limits_{\alpha\in\mathcal{V}_n}p_\rho(\alpha)\Pi_{a,s}A_\alpha\Pi_{a,s}\\
	&=\sum\limits_{\alpha\in\mathcal{V}_n}p_\rho(\alpha)\sum\limits_{\beta\in\mathcal{V}_n}q_{\alpha,a}(\beta,s)A_\beta,
\end{align*}
and so
\begin{equation}\label{eq:quantumState}
	\rho'^{(QM)}=\frac{\sum_\alpha p_\rho(\alpha)\sum_\beta q_{\alpha,a}(\beta,s)A_\beta}{\sum_\alpha p_\rho(\alpha)Q_a(s|\alpha)}.
\end{equation}

Using the classical simulation algorithm, the probability of obtaining outcome $s$ and state $\beta$ given a measurement of $T_a$ on state $\rho$ is $P_{\rho,a}(\beta,s)=P_{\rho,a}(\beta|s)P_{\rho,a}(s)$.  But $P_{\rho,a}(\beta,s)=\sum_\alpha p_\rho(\alpha)P_a(\beta,s|\alpha)=\sum_\alpha p_\rho(\alpha)q_{\alpha,a}(\beta,s)$ and $P_{\rho,a}(\beta|s)=p_{\rho'}(\beta)$.  Therefore, the post-measurement state according to the classical simulation algorithm is
\begin{align}
	\rho'^{(Sim)}&=\sum\limits_{\beta\in\mathcal{V}_n}p_{\rho'}(\beta)A_\beta=\sum\limits_{\beta\in\mathcal{V}_n}\frac{P_{\rho,a}(\beta,s)}{P_{\rho,a(s)}}A_\beta\nonumber\\
	&=\sum\limits_{\beta\in\mathcal{V}_n}\frac{\sum_\alpha p_\rho(\alpha)q_{\alpha,a}(\beta,s)}{\sum_\alpha p_\rho(\alpha)Q_a(s|\alpha)}A_\beta.
\end{align}

This agrees with Eq.~(\ref{eq:quantumState}) above. Therefore, the classical simulation algorithm also reproduces the post-measurement state predicted by quantum mechanics for a single Pauli measurement.

Now let $\rho(t)$ denote the state before the $t\text{th}$ measurement.  Then the above shows that the classical simulation algorithm correctly reproduces the Born rule probabilities $P_{\rho,a_t}(s_t|s_1,s_2,\dots,s_{t-1})$ as well as the post-measurement state $\rho(t+1)$.  Therefore, by induction the simulation algorithm correctly reproduces the outcome probabilities predicted by the Born rule for any sequence of Pauli measurements.
$\Box$\smallskip

{\em{Discussion.}}---We now return to the questions posed in the introduction. 

{\em{(i) Existence of a probability representation.}} It is stated in Theorem~2 of \cite{Ferrie} that ``a quasiprobability representation of quantum theory must have negativity in either its representation of states
or measurements (or both).''

This appears to contradict Theorem~\ref{MT}. However, there is no contradiction here, only a difference in assumptions. Through the definitions made prior to it, the above quoted theorem refers to frame representations. This requires, in particular, that the quasiprobability representation for every quantum state is unique. Clearly, our probability distribution $p$ does not satisfy this condition.

\smallskip

{\em{(ii) Contextuality.}} Given the history of the subject \cite{Howard, Gross,Galv1,Galv2,NegWi,ReWi,Delf2,QuWi,RoM,QuWi19,Heinrich,Kara,Love,Leif,Merm}, an interesting question is whether the present HVM is contextual or noncontextual. The Kochen-Specker notion of contextuality does not apply, because the present value assignments are not deterministic. This leaves us with Spekkens' notion ~\cite{Spekkens2005}, \cite{Ravi} to consider. In this regard, our HVM is preparation contextual and measurement-noncontextual. The former reflects the aforementioned nonuniqueness of $p$.

\smallskip

{\em{(iii) Negativity vs efficiency of simulation.}} Negativity in quasiprobability representations has been identified as a cause for slowing down the classical simulation of quantum systems by sampling. A general result  has been obtained in \cite{Pasha} stating that a quantum system described by a quasiprobality function $W$ with negativity ${\cal{M}}=\|W\|_1$ can be simulated by sampling at a multiplicative cost that scales like ${\cal{M}}^2$.

There are simulation schemes for QCM on qudits \cite{NegWi}, on rebits \cite{ReWi}, and on qubits \cite{QuWi,RoM,QuWi19}, where negativity is the only source for the computational hardness of classical simulation. Negativity is therefore singled out as precondition for quantum speedup.

We do not contradict the results  \cite{NegWi, ReWi, Delf2, QuWi, RoM,QuWi19} but now find that they are an artifact of the particular quasiprobability functions chosen. Our result lies at the opposite end of the spectrum. There is no negativity but, presumably, still computational hardness.

The absence of negativity notwithstanding, there also is continuity with prior works. The probability distribution $p$ satisfies the four criteria of the Stratonovich-Weyl (SW) correspondence~\cite{Strato} (also see~\cite{Brif}; see SM, Section~I for details).  It is thus very closely related to the original Wigner function~\cite{Wig} and to previously defined discrete Wigner functions for finite-dimensional systems.  From the SW perspective, the only condition $p$ doesn't satisfy is uniqueness.

Furthermore, the phase point operators identified in the multiqubit setting of \cite{QuWi19} (also see \cite{KL2}) are special cases of the phase point operators discussed here (see the SM, Section~IV). And thus, the present approach provides a broader and yet conceptually simpler framework for the classical simulation of quantum computation by sampling, subsuming earlier ones as special cases.
\smallskip

{\em{(iv) The PBR theorem.}} The hidden variable model presented here is $\psi$-epistemic~\cite{HarriganSpekkens}.  The PBR theorem~\cite{PBR} asserts that (with certain assumptions) no $\psi$-epistemic model can reproduce the predictions of quantum mechanics.  Our result does not contradict the PBR theorem for two reasons.  First, we consider only sequences of Pauli measurements rather than general measurements (this is sufficent for universal quantum computation).  Second, our model does not satisfy the assumption of preparation independence required for the theorem to hold.  That is, in general, $p_{\rho_1\otimes\rho_2}\ne p_{\rho_1}\cdot p_{\rho_2}$.  

The assumption of preparation independence is less relevant for quantum computation with magic states, where, in the language of resource theories, the free sector is formed by stabilizer states and stabilizer operations, not local states and local operations. Further, the memory lower bound of Karanjai, Wallman, and Bartlett~\cite{Kara} shows that a classical simulation algorithm like that of Algorithm~\ref{Sampling} is incompatible with this assumption.\smallskip

{\em{(v) Gleason's theorem.}} Gleason's theorem \cite{Gleason} says that in Hilbert spaces $H$ of dimension 3 or greater, the only way to assign probabilities $p(h)$ to all subspaces of $h \subset H$, represented by corresponding projectors $\Pi_h$, is via $p(h)=\Tr(\Pi_h \rho)$, for some valid density matrix $\rho$. 

That is, the only consistent way to assign probabilities to measurement outcomes is the Born rule involving density matrices. Our Theorem~\ref{MT} does not contradict this; rather it reproduces the Born rule, cf. Eq.~(\ref{BR}). 

However, Gleason's theorem is sometimes interpreted as a mathematical proof that density operators are the fundamental notion of state in quantum mechanics. In short, density operators are for quantum mechanics, probability distributions for classical statistical mechanics. Theorem~\ref{MT} escapes this interpretation. It shows that every quantum state {\em{can}} be described by a probability distribution, and yet the Born rule for measurement is reproduced. This is possible because we have restricted measurement to Pauli observables. Note though that this restriction does not affect the universality of quantum computation with magic states!\smallskip

To summarize, in this Letter we have constructed a probability function over a finite set capable of positively representing all quantum states on any number of qubits, as well as their update under all Pauli measurements, local and nonlocal. All prior quasiprobability representations invoked in the discussion of quantum computation with magic states, such as the Wigner function for qudits \cite{Gross,NegWi} or for rebits \cite{ReWi}, and the quasiprobability over stabilizer states \cite{RoM}, require negativity to represent universal quantum computation. 

In view of the seeming classicality of the hidden variable model for universal quantum computation constructed here, an important open question is: Where is quantumness hiding?---In this regard, we propose the polytopes $\Lambda_n$, and in particular the algebraic structure of their extremal vertices as a subject for further study.

\medskip

\begin{acknowledgments}
This work is supported by NSERC. We thank Andreas D{\"o}ring (RR) and Bill Unruh (MZ, RR) for discussions.
\end{acknowledgments}

\newpage
\clearpage

\appendix
\setcounter{Theorem}{0}
\setcounter{Cor}{0}
\setcounter{Lemma}{0}
\setcounter{Def}{0}
\setcounter{Conj}{0}
\setcounter{Prop}{0}

$$
\textbf{Supplementary Material}
$$

A comment regarding notation: Equation and Theorem references to the main text carry a suffix ``[m]'' below, to distinguish them from the equation numbering in this supplement. For example, Eq.~(8) from the main text is here referred to as Eq.~(8) [m].

\section{Stratonovich-Weyl correspondence}\label{SW}

In the field of quantum optics, the  Stratonovich-Weyl (SW) correspondence is a set of criteria that well-behaved quasi-probability distributions over phase space have to satisfy.
Denote by $F_A^{(s)}: X \longrightarrow \mathbb{C}$ the quasiprobability distribution corresponding to the (not necessarily Hermitian) operator $A$, with $X$ the phase space and $s$ a real parameter in the interval $[-1,1]$. In the standard formalism for infinite-dimensional Hilbert spaces, $s=-1,0,1$ correspond to the Glauber-Sudarshan $P$, Wigner, and Husimi $Q$ function, respectively. Then, the following set of criteria is imposed on the $F_A^{(s)}$ \cite{StratoSM}; also see  \cite{BrifSM},
\begin{enumerate}
	\item[(0)]{Linearity: $A \longrightarrow F_A^{(s)}$ is a one-to-one linear map.}
	\item[(1)]{Reality: $$F_{A^\dagger}^{(s)}(u) = \left(F_A^{(s)}(u)\right)^*, \forall u \in X.$$}
	\item[(2)]{Standardization: $$\int_X d\mu(u) F_A^{(s)}(u) = \text{Tr}\, A.$$}
	\item[(3)]{Covariance: $$F^{(s)}_{g\cdot A}(u) = F^{(s)}_A(g^{-1} u), \; g\in G,$$ with $G$ the dynamical symmetry group.}
	\item[(4)]{Traciality: $$\int_X d\mu(u) F_A^{(s)}(u)F_B^{(-s)}(u) = \text{Tr}\, AB.$$}
\end{enumerate}
To investigate the SW criteria in the present setting, we first extend the probability distributions $p_\rho$ defined in Eq.~(2) [m] for proper density matrices to a quasiprobability function $W$ defined for all   operators $A$, via
\begin{equation}\label{ext}
A = \sum_\alpha W_A(\alpha) A_\alpha.
\end{equation}
We note that $W$ does not come with a parameter $s$; there is only a single quasiprobability function $W$. This will affect the formulation of traciality. 

Further, the mapping $A \longrightarrow W_A$ is linear, $A+B$ can be represented as $W_A+W_B$. However, the mapping is one-to-many, and the Stratonovich-Weyl criterion (0) is thus not satisfied. In fact, this is a general consequence of Kochen-Specker contextuality, as has been demonstrated in \cite{LeiferMaroney2013SM}.

The remaining SW conditions apply. 

{\em{(1) Reality.}} All phase point operators $A_\alpha$ are Hermitian by definition, cf. Eq.~(1) [m].
Therefore $A^\dagger$ can be represented by the quasiprobability distribution $\alpha\mapsto W_A(\alpha)^*$.

{\em{(2) Standardization.}} By their definition Eq.~(1) [m], the phase point operators satisfy $\text{Tr}\, A_\alpha=1$, for all $\alpha \in \mathcal{V}_n$. Standardization,
\begin{equation}\label{Stand}
\text{Tr} A = \sum_\alpha W_A(\alpha),
\end{equation} 
follows by taking the trace of Eq.~(\ref{ext}). 

{\em{(3) Covariance.}} Let $\text{Cl}_n$ denote the $n$-qubit Clifford group.
We have the following result.
\begin{Lemma}\label{Cov}
	For any operator $A$  it holds that
	\begin{equation}\label{Cov1}
	W_{gAg^\dagger}(\alpha) = W_A(g^{-1}\alpha),\;\;\forall g\in \text{Cl}_n.
	\end{equation}
\end{Lemma}
{\em{Proof of Lemma~\ref{Cov}.}} First we show that $\Lambda_n$ is mapped into itself under the action of the Clifford group. Namely, for all stabilizer sates $|\sigma\rangle \in {\cal{S}}$,
$$
\begin{array}{rcl}
\text{Tr}(gA_\alpha g^\dagger |\sigma\rangle \langle \sigma|) &=& \text{Tr}(A_\alpha \, g^\dagger |\sigma\rangle \langle \sigma| g)\\
&=& \text{Tr}(A_\alpha \, |\sigma'\rangle \langle \sigma'|)\\
&\geq &0.
\end{array}
$$
Furthermore, $\text{Tr}(gA_\alpha g^\dagger)=\text{Tr} A_\alpha =1$. Hence, with the definition Eq.~(1) [m] of $\Lambda_n$, it holds that $gA_\alpha g^\dagger \in \Lambda_n$, for all $\alpha \in \mathcal{V}_n$ and all $g \in \text{Cl}_n$.

Now we show that for every $\alpha \in \mathcal{V}_n$ and every $g\in \text{Cl}_n$ there is a unique $\beta \in \mathcal{V}_n$ such that 
\begin{equation}\label{albet}
gA_\alpha g^\dagger = A_\beta.
\end{equation}
Let $\mathcal{S}_\alpha$ be the subset of stabilizer states that specifies $A_\alpha$, i.e. $A_\alpha$ is the unique solution in $\Lambda_n$ to the set of constraints $\Tr(X|\sigma\rangle\langle\sigma|)=0$ for all $|\sigma\rangle\in \mathcal{S}_\alpha$. In fact, we can choose the size of $\mathcal{S}_\alpha$ to be equal to $2^{2n}-1$ \cite[Theorem 18.1]{LPSM}.
Let $g\cdot S_\alpha$ denote the set of stabilizers $g|\sigma\rangle\langle\sigma|g^\dagger$ where $|\sigma\rangle\in \mathcal{S}_\alpha$. Then the action of $g$ gives a one-to-one correspondence between the set of solutions to the constraints specified by $\mathcal{S}_\alpha$ and $g^\dagger\cdot \mathcal{S}_\alpha$ since if $X$ is a solution to the former then $g X g^\dagger$ is a solution to the latter and vice versa. Moreover, $gXg^\dagger$ belongs to the polytope $\Lambda_n$. Therefore $g A_\alpha g^\dagger$ specifies a vertex.  In other words, given $\alpha \in \mathcal{V}_n$ and $g \in \text{Cl}_n$, Eq.~(\ref{albet}) holds for a suitable $\beta \in \mathcal{V}_n$. We thus define $g\alpha:=\beta$, and Eq.~(\ref{albet}) becomes
\begin{equation}\label{albet2}
g A_\alpha g^\dagger = A_{g \alpha}.
\end{equation}
Therefore,
$$
\begin{array}{rcl}
\sum_\alpha W_{gAg^\dagger}(\alpha) A_\alpha &=& gAg^\dagger\\ &=& \sum_\alpha W_A(\alpha) \, gA_\alpha g^\dagger\\ 
&=& \sum_\alpha W_A(\alpha) A_{g\alpha}\\
&=& \sum_\alpha W_A(g^{-1} \alpha) A_{\alpha}.
\end{array}
$$
Thus, $W_{gAg^\dagger}(\alpha) = W_A(g^{-1} \alpha) A_{\alpha}$. $\Box$

We remark that, for qubits, only non-unique quasi\-probability functions can be Clifford covariant. Namely, if the phase point operators form an operator basis, i.e., are linearly independent, then the resulting quasiprobability function cannot be Clifford covariant \cite{ZhuSM}.

The covariance property can be used to efficiently simulate the effect of Clifford unitaries in QCM, as an alternative to the method of treating Clifford unitaries discussed in the main text.

{\em{(4) Traciality.}} In the absence of a continuously varying parameter $s$, we introduce a dual quasiprobability function $\tilde{W}$ in addition to $W$, to stand in for $F^{(-s)}$. Namely, for all projectors $\Pi_{a,s}$,  corresponding to measurements of Pauli observables $T_a$ with outcome $s$, we define
$$
\tilde{W}_{\Pi_{a,s}}(\alpha):= Q_a(s|\alpha).
$$
By linearity, this implies expressions for all $\tilde{W}_{T_a}(\alpha)$. Since the Pauli operators form an operator basis, again by linearity one obtains $\tilde{W}_A$ for any operator $A$. Then,
$$
\text{Tr} AB = \sum_\alpha \tilde{W}_A(\alpha) W_B(\alpha)
$$
follows from Eq.~(4) [m].

We thus satisfy the SW criteria (1) - (4).\smallskip

To conclude, we emphasize that for the present purpose of classically simulating QCM, a crucial property of $W$ is positivity preservation under Pauli measurement. This property has no counterpart in the Stratonovich-Weyl correspondence.

\section{Some background on QCM}

Quantum computation with magic states (QCM) is a scheme for universal quantum computation, closely related to the circuit model. From a practical point of view, QCM is very advantageous for fault-tolerant quantum computation \cite{BKSM}, but that does not concern us here.

\subsection{Operations in QCM}

There are two types of operations in QCM, the ``free'' operations and the resources. The free operations are (i) Preparation of all stabilizer states, (ii) All Clifford unitaries, and (iii) Measurement of all Pauli observables.

The resource are arbitrarily many copies of the state
\begin{equation}\label{MagStat}
|{\cal{T}}\rangle = \frac{|0\rangle + e^{i\pi/4}|1\rangle}{\sqrt{2}}.
\end{equation}
The state $|{\cal{T}}\rangle$ is called a ``magic state''.

A stabilizer state is a pure $n$-qubit quantum state which is the joint eigenstate of a maximal set of commuting Pauli operators \cite{GomaSM, Goma2SM, NieChaSM}. The $n$-qubit Clifford group $Cl_n$ is the largest subgroup of $SU(2^n)$ with the property that for any $g\in Cl_n$ and all Pauli operators $T_a$ there exists a Pauli operator $T_b$ such that
$$
gT_ag^\dagger = \pm T_b.
$$
That is, the Clifford group is the normalizer of the Pauli group.\smallskip

The distinction between free operations and resources in QCM is motivated by the Gottesman-Knill theorem. Namely, the free operations alone are not universal for quantum computation, and, in fact, can be efficiently classically simulated. The magic states restore computational universality (see below), hence the name.

A further motivation for subdividing the computational primitives  into free operations and resources stems from quantum error correction. Fault-tolerant versions of the free operations are comparatively easy to produce, but the creation of fault-tolerant magic states  is very operationally costly.

\subsection{Computational universality}

It is well known \cite{YaoSM} that the gates
$$
\{\text{CNOT}_{ij}, H_i, {\cal{T}}_i,\; 1\leq i\neq j \leq n\}
$$
form a universal set, i.e., enable universal quantum computation. Therein, the controlled NOT gates $\text{CNOT}_{ij}$  between qubits $i$ and $j$ and the Hadamard gates $H_i$ are in the Clifford group. The only non-Clifford element in the above universal set is
$$
{\cal{T}}_i = \exp \left(-i\frac{\pi}{8} Z_i\right).
$$
This gate can be simulated by the use of a single magic state $|{\cal{T}}\rangle$ in a circuit of Clifford gates and Pauli measurements (circuit reproduced from Fig. 10.25 of \cite{NieChaSM}),
$$
\parbox{5.1cm}{\includegraphics[width=5cm]{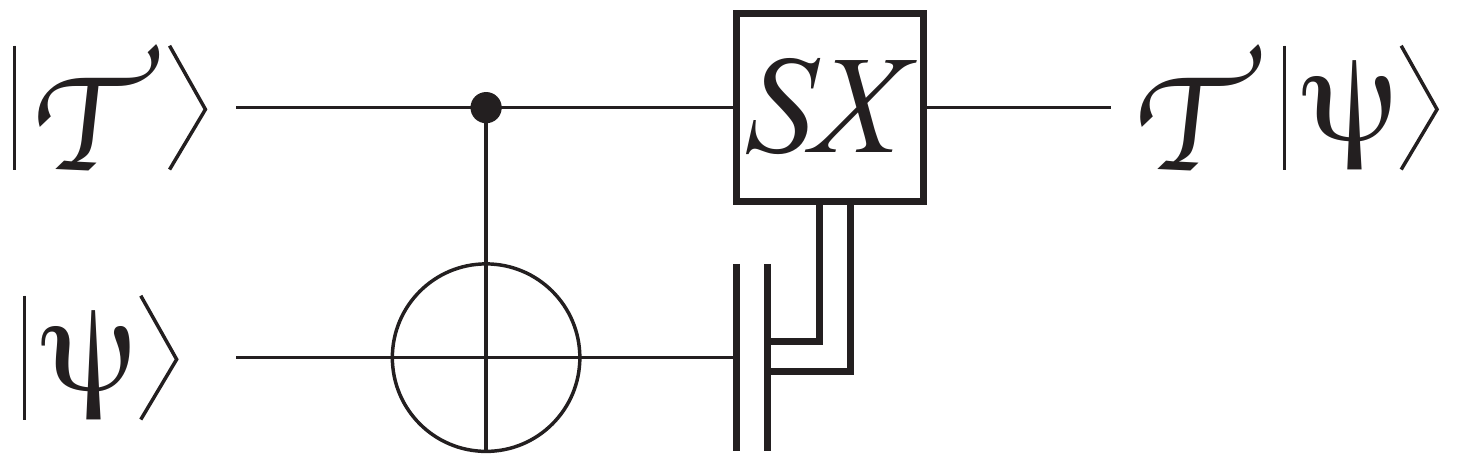}}.
$$ 
Therein, the lower qubit is measured in the $Z$-basis, and the binary measurement outcome classically controls the $SX$-gate. $S$ is a Clifford gate,
$$
S_i=\exp \left(-i\frac{\pi}{4} Z_i\right).
$$
Thus, the magic states Eq.~(\ref{MagStat}) boost the free operations to quantum computational universality.

\section{Complexity parameter of the state polytope $\Lambda_n$}

A question that arises with Theorem~2 [m] is what determines the value of $n$ labelling the state polytope $\Lambda_n$, and hence the complexity of the simulation. In this regard, we make the following observation.
\begin{Lemma}\label{L0}
	Any quantum computation in the magic state model that operates on an initial state $|\mu\rangle_A \otimes |\sigma\rangle_B$, where $|\mu\rangle$ is an $n$-qubit magic state and $|\sigma\rangle$ is an $m$-qubit stabilizer state, can with the same efficiency be run on the magic state $|\mu\rangle$ alone.  
\end{Lemma}
Supplementing the non-stabilizer magic state $|\mu\rangle$ with stabilizer states is thus redundant. For example, if the magic states used in a given QCM are all of $T$-type, then $n$ can be taken to be the number of those states.
\smallskip

{\em{Proof of Lemma~\ref{L0}.}} Wlog. we discuss the version of QCM where the quantum computation consists of a sequence of only Pauli measurements. We give an explicit procedure to replace the sequence $\tau$ on $A \otimes B$ by an equivalent sequence $\tilde\tau^{(A)}$ of measured observables that act only on the subsystem $A$. The proof is by induction, and the induction hypothesis is that, at time $t$, the sequence $\tau_{\leq t}$ of measurements has been replaced by a computationally equivalent sequence $\tilde\tau^{(A)}_{\leq t}$ of Pauli measurements on the register $A$ only. This statement is true for $t=0$, i.e., the empty measurement sequence. We now show that the above statement for time $t$ implies the analogous statement for time $t+1$.

At time $t$, the state of the quantum register evolved under the computationally equivalent measurement sequence $\tilde\tau^{(A)}_{\leq t}$ is $|\Psi(t)\rangle = |\psi(t)\rangle_A\otimes |\sigma\rangle_B$. We now consider the Pauli observable $T(t+1)\in \tau$ to be measured next, and write $T(t+1)= S_A(t+1) \otimes R_B(t+1)$. There are two cases.  \smallskip

Case I: {\em{$T(t+1)$ commutes with the entire stabilizer ${\cal{S}}$ of $|\sigma\rangle$.}} Hence, also $R_B(t+1)$ commutes with ${\cal{S}}$. But then, either $R_B(t+1)$ or $-R_B(t+1)$ is in ${\cal{S}}$, and $R_B(t+1)$ may be replaced by its eigenvalue $\pm 1$ in the measurement. Hence, the measurement of $T(t+1)$ is equivalent to the measurement of $\pm S_A(t+1)$.

Case II: {\em{$T(t+1)$ does not commute with the entire stabilizer ${\cal{S}}$ of $|\sigma\rangle$.}} Then, the measurement outcome $s_{t+1}$ is completely random. Further, there exists a Clifford unitary $U$ such that
$$ 
\begin{array}{rcl}
U {\cal{S}} U^\dagger &=& \langle X_{B:1}, X_{B:2},..,X_{B:m}\rangle,\\
U T(t+1) U^\dagger &=& Z_{B:1}.
\end{array}
$$
Therefore, the state resulting from the measurement of $T(t+1)$, with outcome $s_{t+1}$ on the state $|\Psi(t)\rangle$ is the same state as the one resulting from the following procedure: 
\begin{enumerate}
	\item{Apply the Clifford unitary $U$ to $|\Psi(t)\rangle  = |\psi(t)\rangle_A\otimes |\sigma(t)\rangle_B$, leading to
		$$
		U\, |\Psi(t)\rangle = |\tilde{\psi}(t)\rangle \otimes |\overline{+}\rangle_B,
		$$
		where $|\overline{+}\rangle_B:=\bigotimes_{i\in B}|+\rangle_{B:i}$.}
	\item{Measure $Z_{B:1}$ on $ |\tilde{\psi}(t)\rangle \otimes |\overline{+}\rangle_B$, with outcome $s_{t+1}$.}
	\item{Apply $U^\dagger$.}
\end{enumerate}
Now, note that the measurement in Step 2, of the Pauli observable $Z_{B:1}$ is applied to the stabilizer state $|\overline{+}\rangle_B$. The result is $|\tilde{\sigma}(t+1)\rangle = |s_{t+1}\rangle_{B:1} \bigotimes_{j=2}^m |+\rangle_{B:j}$. Therefore, after normalization, the effect of the measurement can be replaced by the unitary $\left(X_{B:1}\right)^{s_{t+1}} H_{B:1}$.

Thus, the whole procedure may be replaced by the Clifford unitary $U^\dagger \, \left(X_{B:1}\right)^{s_{t+1}} H_{B:1} \, U$. But Clifford unitaries don't need to be implemented. They are just propagated past the last measurement, thereby affecting the measured observables by conjugation whereby their Pauli-ness is preserved. In result, in Case II, the measurement of $T(t+1)$ doesn't need to be performed at all. It is replaced by classical post-processing of the  subsequent measurement sequence.\medskip

We conclude that in both the cases I and II, given the induction assumption, the original measurement sequence $\tau_{\leq t+1}$ can be replaced by a computationally equivalent measurement sequence $\tilde\tau^{(A)}_{\leq t+1}$ acting on register $A$ only. By induction, the complete measurement sequence $\tau$ can be replaced by a computationally equivalent sequence $\tilde\tau^{(A)}$ acting on $A$ only.

Since the measurements $\tilde\tau^{(A)}$ are applied to an unentangled initial state $|\mu\rangle_A \otimes |\sigma\rangle_B$, the register $B$ can be dropped. Finally, the measurement sequence  $\tilde\tau^{(A)}$ is of the same length or shorter than $\tau$, and can be efficiently computed from the latter. Hence its implementation is at least as efficient. $\Box$

\section{Multi-qubit phase points from \cite{QuWi19SM} are extremal}\label{Extr}

The present work, there is no negativity anywhere in the classical simulation of QCM. The shifting of the cause for computational hardness away from negativity to other potential sources is a major disruption with the prior works \cite{NegWiSM,Delf2SM,ReWiSM,QuWiSM,QuWi19SM}. 

But underneath this discontinuity lies an element of continuity. Namely, the direct precursor to the present work is Ref.~\cite{QuWi19SM}; and the phase point operators of the multi-qubit quasiproability function defined therein are also extremal vertices of the present state polytope $\Lambda_n$. This is the content of Lemma~\ref{extrem} below, the main result of this section. It shows that the multi-qubit phase space defined in \cite{QuWi19SM} is a subset of the phase space of the present model, describing a sector of it in which the update rules under Clifford unitaries and Pauli measurements are guaranteed to be computationally efficient.\medskip

Recall from \cite{QuWi19SM} a couple of definitions. We call a set $\Omega \subset E_n$ closed under inference if for all $a,b \in \Omega$ with the property that $[a,b]=0$ it holds that $a+b \in \Omega$. (Here $[a,b]:= a_X b_Z + a_Z b_X\mod 2$.)
We call a set $\Omega \subset E_n$ non-contextual if it supports a non-contextual value assignment. Sets $\Omega$ which are both closed under inference and non-contextual are called ``cnc'' \cite{QuWi19SM} (also see \cite{KL2SM}). Of particular interest in are maximal cnc sets, which are cnc sets that are not strictly contained in any other cnc set. They give rise to the following multi-qubit phase point operators
\begin{equation}\label{PhaPoi}
A_\Omega^\gamma=\frac{1}{2^n}\sum_{a\in \Omega} (-1)^{\gamma(a)}T_a,
\end{equation}
where $\Omega$ is a maximal cnc set, and $\gamma: \Omega \longrightarrow \mathbb{Z}_2$ is a non-contextual value assignment. 

Theorem~1 in \cite{QuWi19SM} classifies the maximal cnc sets. For the present purpose it may be rephrased as
\begin{Lemma}\label{Classif}
	If a subset of $E_n$ is closed under inference and does not contain a Mermin square then it is non-contextual.
\end{Lemma}
{\em{Proof sketch for Lemma~\ref{Classif}.}} Theorem~1 of \cite{QuWi19SM} classifies the subsets of $E_n$ that are closed under inference and do not contain a Mermin square. They all turn out to be non-contextual. $\Box$\smallskip

We now have the following result (also see \cite{HeimSM} for an independent proof).
\begin{Lemma}\label{extrem}
	For any number $n$ of qubits, the phase point operators $A_\Omega^\gamma$ of Eq.~(\ref{PhaPoi}) are vertices of $\Lambda_n$.
\end{Lemma}

An independent proof of this result is given in~\cite{HeimSM}.

{\em{Proof of Lemma~\ref{extrem}.}} Pick an $n$, any pair $(\Omega,\gamma)$. $A_\Omega^\gamma$ has unit trace, and, as shown in \cite{QuWi19SM}, satisfies $\text{Tr}(|\sigma\rangle \langle \sigma| A_\Omega^\gamma)\geq 0$. Therefore, $A_\Omega^\gamma \in \Lambda_n$, and $A_\Omega^\gamma$ has an expansion
\begin{equation}\label{exp2}
A_\Omega^\gamma = \sum_{\beta\in\mathcal{V}_n} p_{\Omega,\gamma}(\beta)A_\beta,
\end{equation}
where $p_{\Omega,\gamma}(\beta)\geq 0$, $\forall \beta$, and $\sum_\beta p_{\Omega,\gamma}(\beta) =1$. Thus, $p_{\Omega,\beta}$ is a probability distribution. Henceforth, we consider any $A_\beta$ for which $p_{\Omega,\gamma}(\beta)>0$. 

Now pick an $a\in \Omega$ and consider $\text{Tr}\left(T_a A_\Omega^\gamma\right)$. With Eq.~(\ref{PhaPoi}), it holds that $(-1)^{\gamma(a)} = \sum_\beta p_{\Omega,\gamma}(\beta) \langle T_a\rangle_\beta$. Since $p_{\Omega,\beta}$ is a probability distribution and $|\langle T_a\rangle_\beta|\leq 1$ for all $\beta$, it follows that
$$
\langle T_a\rangle_\beta =(-1)^{\gamma(a)},\;\forall \,\beta\;\text{with}\; p_{\Omega,\gamma}(\beta)>0.
$$
That is, every phase point operator that appears on the rhs. of Eq.~(\ref{exp2}) with non-zero coefficient agrees with $A_\Omega^\gamma$ on the expectation values $\langle T_a\rangle$ for all $a\in \Omega$.

Now we turn to the expectation values for $b \not\in \Omega$. 
Any set $\tilde{\Omega} \subset E_n$ that is closed under inference and contains both $\Omega$ and $b$ is contextual, by the maximality of $\Omega$. By Lemma~\ref{Classif}, any such $\tilde{\Omega}$ contains a Mermin square $M$, and furthermore $b\in M$. 

\begin{figure}
	\begin{center}
		\begin{tabular}{lclcl}
			(a) && (b)\\
			\includegraphics[width=2.5cm]{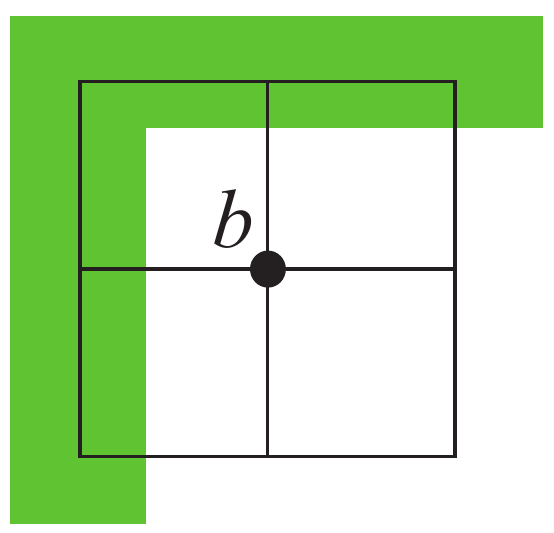} && \includegraphics[width=2.5cm]{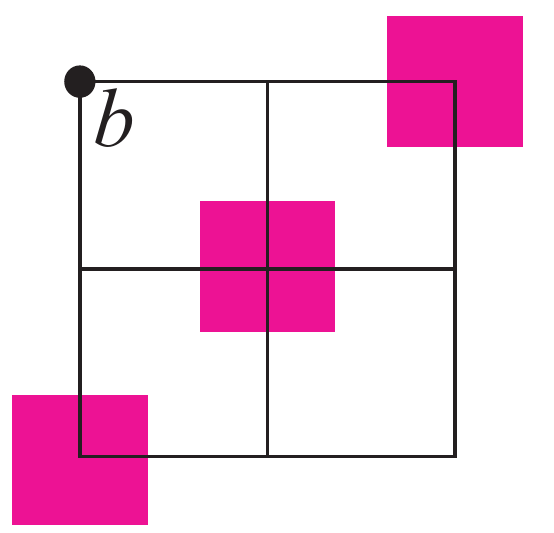}
		\end{tabular}
		\caption{\label{OM}Two possibilities for the set $\Omega \cap M$, shown in color.}
	\end{center}
\end{figure}

Since $M$ is closed under inference, so is $\Omega \cap M$. Also, since $\Omega$ is maximal, $\Omega \cap M$ is maximal in $M$. Up to permutations of rows and columns, there are two possibilities for $\Omega \cap M$, which are displayed in Fig.~\ref{OM}. 

{\em{Case (a).}} For any $b$ there exists a triple $\{x,y,z\} \subset M\backslash b$ such that $[x,y]=[x,z]=[b,y]=[b,z]=0$, $[x,b]=[y,z]\neq 0$. We have the following Mermin square:
$$
\includegraphics[width=3.3cm]{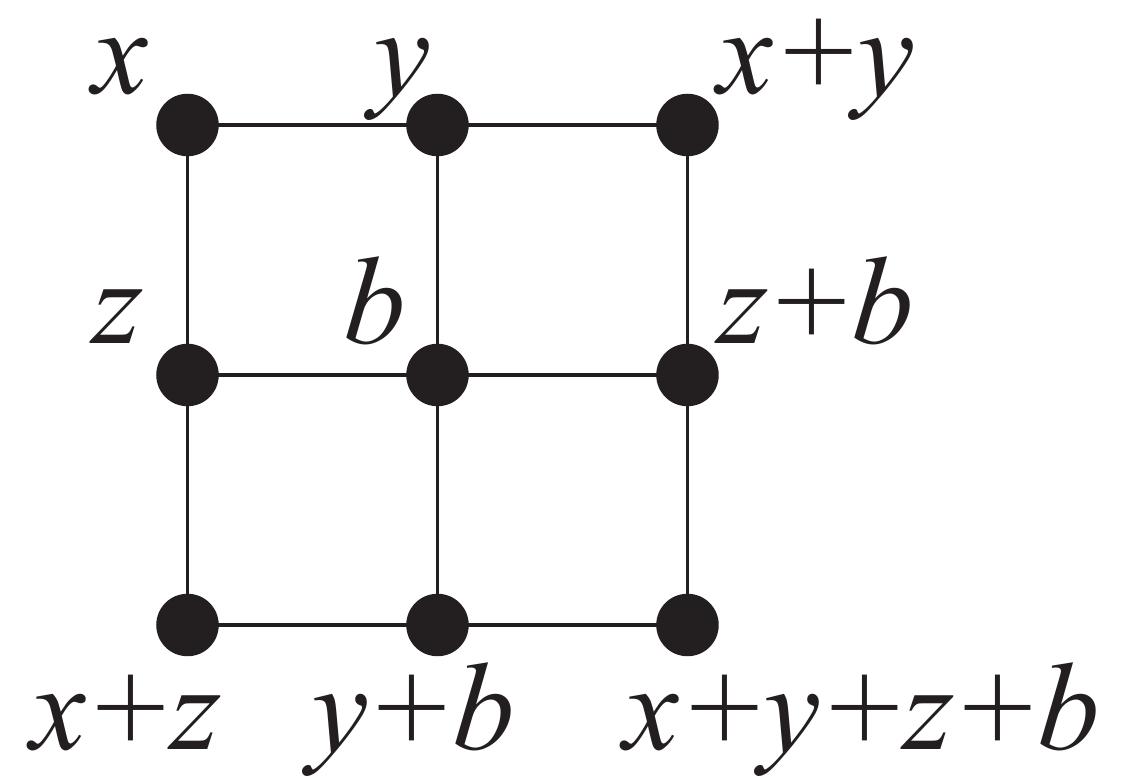}
$$
Therein, Mermin's contradiction to the existence of a non-contextual HVM is encapsulated in the operator relation $(T_xT_y)(T_zT_b) = - (T_xT_z)(T_yT_b)$.

We chose the following phase conventions.
\begin{equation}
\label{phasconA}
\begin{array}{rcl}
T_{x+y}=T_xT_y,\; T_{z+b}=T_zT_b,\\
T_{x+z}=T_xT_z,\;T_{y+b}=T_yT_b,
\end{array}
\end{equation}
and
\begin{equation}
\label{phasconB}
\begin{array}{rcl}
T_{x+y+z+b}&=&T_{x+z}T_{y+b},\\ 
T_{x+y+z+b}&=&-T_{x+y}T_{z+b}.
\end{array}
\end{equation}
Recall that with the first part of the proof $\langle T_j\rangle_\beta=(-1)^{\gamma(j)}$, for $j = x,y,z$. Now assume that $\langle T_b\rangle_\beta = \nu$, with $-1\leq \nu \leq 1$. Now, with Eq.~(\ref{phasconA})
$$
\begin{array}{rcl}
\langle T_{x+y} \rangle_\beta &=& (-1)^{\gamma(x)+\gamma(y)},\; \langle T_{x+z} \rangle_\beta = (-1)^{\gamma(x)+\gamma(z)},\\
\; \langle T_{y+b} \rangle_\beta &=& \nu(-1)^{\gamma(y)},\;\langle T_{z+b} \rangle_\beta = \nu(-1)^{\gamma(z)}.
\end{array}
$$
Therefore, with Eq.~(\ref{phasconB}),
$$
\begin{array}{rcl}
\langle T_{x+y+z+b}\rangle_\beta &=&\nu (-1)^{\gamma(x)+\gamma(y)+\gamma(z)} \\ 
&=& -\nu (-1)^{\gamma(x)+\gamma(y)+\gamma(z)}.
\end{array}
$$
This is satisfiable only if $\nu=0$, and hence $\langle T_b\rangle_\beta=0$. 

{\em{Case (b).}} The argument is analogous to case (a), and we do not repeat it here.

By the above case distinction, for any $b \in E_n\backslash \Omega$ either case (a) or (b) applies, and each way the consequence is that $\langle T_b\rangle_\beta =0$. Therefore, any phase point operator $A_\beta$ that appears on the rhs of Eq.~(\ref{exp2}) with nonzero $p_{\Omega,\gamma}(\beta)$ agrees with $A_\Omega^\gamma$ on {\em{all}} expectation values of Pauli observables; hence $A_\Omega^\gamma = A_\beta$ for all such $\beta$.

Now assume there exists no such $A_\beta$. Taking the trace of Eq.~(\ref{exp2}) yields $1=0$; contradiction. Hence, there must exist a $\beta$ such that $A_\Omega^\gamma =A_\beta$, for all $(\Omega,\gamma)$. $\Box$

\section{The two-qubit polytope $\Lambda_2$}\label{twoQubitLambda}
\begin{figure}
	\centering
	\includegraphics[width=8.6cm]{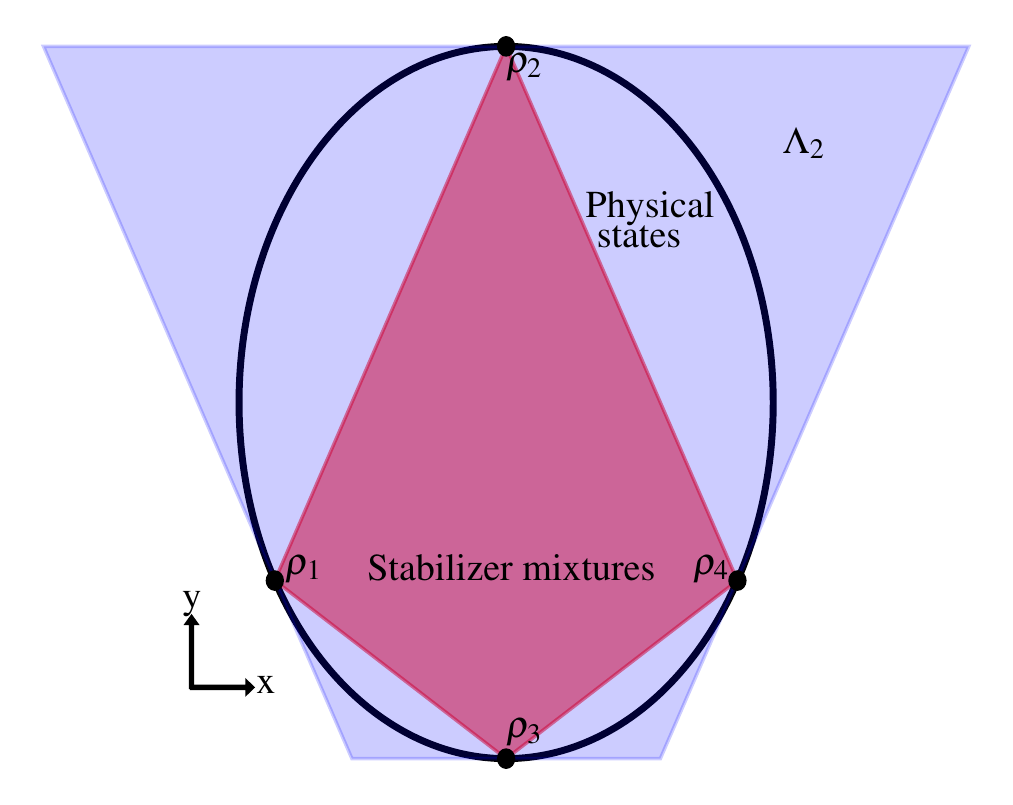}
	\caption{\label{fig:twoQubitCrossSection}Cross section of the space $\text{Herm}_1(4)$ parameterized by Eq.~\ref{eq:twoQubitCrossSection}.  The two-qubit stabilizer polytope is inscribed in the set of physical states and the set of physical states is inscribed in the polytope $\Lambda_2$.  The states labelled by $\rho_1$--$\rho_4$ are given in Eq.~(\ref{eq:twoQubitCrossSectionStates}).}
\end{figure}

Fig.~1 in the main text shows what the polytope $\Lambda_n$ looks like for a single qubit, $n=1$.  The polytope $\Lambda_1$ is a cube inscribing the Bloch ball---the set of physical quantum states.  The situation is similar for multiple qubits.  In general, $\Lambda_n$ is not a hypercube, it is a more general polytope, but it still inscribes the set of physical states.

Fig.~\ref{fig:twoQubitCrossSection} shows a cross section of the space $\text{Herm}_1(4)$, indicating the states which are contained in the two-qubit stabilizer polytope---the set of mixtures of pure two-qubit stabilizer states, the set of physical states, and the polytope $\Lambda_2$.  The cross section is parameterized by
\begin{equation}\label{eq:twoQubitCrossSection}
\rho(x,y)=\frac{1}{4}I_{12}+x(Z_1+Z_2)+y(X_1X_2+Z_1Z_2-Y_1Y_2).
\end{equation}
The four states labelled in the figure are
\begin{equation}\label{eq:twoQubitCrossSectionStates}
\begin{array}{rcl} 
\rho_1&=& \displaystyle{\frac{1}{4}I_{12}-\frac{1}{8}(Z_1+Z_2)},\vspace{1mm}\\
\rho_2&=& \displaystyle{\frac{1}{4}I_{12}+\frac{1}{4}(X_1X_2+Z_1Z_2-Y_1Y_2)}\vspace{1mm},\\
\rho_3&=& \displaystyle{\frac{1}{4}I_{12}-\frac{1}{12}(X_1X_2+Z_1Z_2-Y_1Y_2)}\vspace{1mm},\\
\rho_4&=&  \displaystyle{\frac{1}{4}I_{12}+\frac{1}{8}(Z_1+Z_2)}.
\end{array}	
\end{equation}

\section{$\Lambda_n$ is bounded}\label{Lambda-bounded}
The set $E_n$ has the structure of a vector space over $\mathbb{Z}_2$. The commutator $T_a T_b T_aT_b$ is given by $(-1)^{[a,b]}$ where $[a,b]=a_Z^T b_X+b_Z^T a_X \mod 2$. A subspace of $E_n$ on which the symplectic form $[\cdot,\cdot]$ vanishes is called an isotropic subspace. For an isotropic subspace $I\subset E_n$ and a value assignment $\lambda:I\to \mathbb{Z}_2$ we define a projector
$$
\Pi_{I,\lambda} = \frac{1}{|I|} \sum_{a\in I} (-1)^{\lambda(a)} T_a.
$$
Summing over all value assignments gives a resolution of the identity: $\sum_{\lambda} \Pi_{I,\lambda} = \mathbb{I}$.
For each stabilizer state $|\sigma\rangle$ there is a unique pair $(I,\lambda)$ consisting of a maximal isotropic subspace and a value assignment defined on it such that $\Pi_{I,\lambda} = |\sigma\rangle \langle \sigma|$. Then for $X\in \Lambda_n$ we have
$$
\begin{array}{rcl}
\Tr(X \Pi_{a,s}) &=& \Tr(X\Pi_{a,s} \mathbb{I}) \\
&=& \Tr(X\Pi_{a,s} \sum_{\lambda'} \Pi_{I',\lambda'}) \\
&=& \sum_{\lambda'} \Tr(X\Pi_{a,s}  \Pi_{I',\lambda'}) \\
&=&\sum_{\lambda|\;\lambda(a)=s} \Tr(X \Pi_{I,\lambda}) \geq 0.
\end{array}
$$
Therefore $\Lambda_n$ is contained in the hypercube defined by
$$\lbrace X\in \text{Herm}_1(2^n)|\; \Tr(\Pi_{a,s} X)\geq 0,\; \forall a\in E_n-\lbrace 0 \rbrace,\; s=0,1 \rbrace$$
and thus it is bounded.


\begin{thebibliography}{99}

\bibitem{QTCM}
A. Peres, {\em{Quantum Theory: Concepts and Methods}}, Springer, 1995.

\bibitem{Fey}
R.P. Feynman, R.B. Leighton, M. Sands, {\em{Probability Amplitudes}}. The Feynman Lectures on Physics. Volume 3. Redwood City: Addison-Wesley (1989). 

\bibitem{BK}
S. Bravyi and A. Kitaev, Phys. Rev. A \textbf{71}, 022316 (2005).


\bibitem{Bartl2}
J.J. Wallman, S.D. Bartlett, Phys. Rev. A \textbf{85}, 062121 (2012).

\bibitem{Pasha}
H. Pashayan, J.J. Wallman, S.D. Bartlett, Phys. Rev. Lett. \textbf{115}, 070501 (2015).

\bibitem{Gross}
D. Gross, {\em{Computational Power of Quantum Many-Body States and Some Results on Discrete Phase Spaces}}, Ph.D thesis, Imperial College London, 2005.

\bibitem{Galv1}
E.F. Galv{\~a}o, Phys. Rev. A \textbf{71}, 042302 (2005).

\bibitem{Galv2}
C. Cormick, E.F. Galv{\~a}o, D. Gottesman, J.P. Paz, and A.O. Pittenger,  Phys. Rev A \textbf{73}, 012301 (2006).

\bibitem{NegWi}
V. Veitch, C. Ferrie, D. Gross, and J. Emerson, New J. Phys. \textbf{14}, 113011 (2012).

\bibitem{ReWi} 
N. Delfosse, P. Allard Guerin, J. Bian and R. Raussendorf, Phys. Rev. X \textbf{5}, 021003 (2015).

\bibitem{Delf2}
N. Delfosse, C. Okay, J. Bermejo-Vega, D. E. Browne, and R. Raussendorf, New J. Phys. \textbf{19}, 123024 (2017).

\bibitem{QuWi}
R. Raussendorf, D. E. Browne, N. Delfosse, C. Okay, and J. Bermejo-Vega, Phys. Rev. A \textbf{95}, 052334 (2017).

\bibitem{QuWi19}
R. Raussendorf, J. Bermejo-Vega, E. Tyhurst, C. Okay, M. Zurel, Phys. Rev. A \textbf{101}, 012350 (2020).

\bibitem{RoM}
M. Howard, E.T. Campbell, Phys. Rev. Lett. \textbf{118}, 090501 (2017).

\bibitem{Heinrich}
M. Heinrich, D. Gross, Quantum \textbf{3}, 132 (2019).

\bibitem{Kara}
A. Karanjai, J.J. Wallman, S.D. Bartlett, arXiv:\-1802.\-07744v1.

\bibitem{Zhu}
H. Zhu, Phys. Rev. Lett. \textbf{116}, 040501 (2016).

\bibitem{Love}
L. Kocia and P. Love, Phys. Rev. A \textbf{96}, 062134 (2017).

\bibitem{Brot}
J.B. DeBrota, C.A. Fuchs, and B.C. Stacey, Phys. Rev. Research 2, 013074 (2020).

\bibitem{Brot2}
J.B. DeBrota, B.C. Stacey, arXiv:1912.07554

\bibitem{Hudson1974}
R. L. Hudson, Rep. Math. Phys. \textbf{6}, 249-252
(1974).

\bibitem{KenfackZyczkowski2004}
A. Kenfack and K. {\.Z}yczkowski, J. Opt. B \textbf{6}, 396-404 (2004).

\bibitem{Ferrie}
C. Ferrie,  Rep. Prog. Phys. 74, 115001 (2011).

\bibitem{PBR}
M.~Pusey, J.~Barrett, T.~Rudolph, Nat.~Phys.~\textbf{8}, 475-478 (2012).

\bibitem{Goma}
D. Gottesman, Proceedings of the XXII International Colloquium on Group Theoretical Methods in Physics, p. 32-43 (Cambridge, MA, International Press, 1999)

\bibitem{Goma2}
S. Aaronson, D. Gottesman, Phys. Rev. A \textbf{70}, 052328 (2004).

\bibitem{NieCha}
M.A. Nielsen and I.L. Chuang, {\em{Quantum Computation and Quantum Information}},
Cambridge University Press (2000).

\bibitem{Heim}
A. Heimendahl, MSc Thesis, University of Cologne, 2019.

\bibitem{Ziegler1995}
G. M. Ziegler, Lectures on Polytopes, Springer-Verlag, New York, (1995).

\bibitem{LP}
V. Chvatal, Linear programming, W. H. Freeman and Company, New York,  (1983).

\bibitem{EPR}
A. Einstein, B. Podolsky, and N. Rosen, Phys. Rev. \textbf{47}, 777 (1935).

\bibitem{Bell}
J. Bell, Physics \textbf{1}, 195 (1964).

\bibitem{KS}
S. Kochen and E. P. Specker,  J. Mathematics and Mechanics \textbf{17}, 59 (1967).

\bibitem{Bohm}
D. Bohm, Phys. Rev. \textbf{85}, 166 (1952).

\bibitem{BB}
E.G. Beltrametti S. Bugajski,  J. Phys. A: Math. Gen. \textbf{29}, 247 (1996).

\bibitem{Yao}
A. Yao, Proc. 34th Annual IEEE Symposium on Foundations of Computer Science, 352--361 (1993).

\bibitem{Vidal2003}
G. Vidal, Phys. Rev. Lett. \textbf{91}, 147902 (2003).

\bibitem{Shep}
M.J. Bremner , R. Jozsa  and D.J. Shepherd, Proc. R. Soc. A,  \textbf{467}, 459 (2011).

\bibitem{Howard}
M. Howard {\em{et al.}}, Nature \textbf{510}, 351355 (2014).

\bibitem{Leif}
M.S. Leifer and O.J.E. Maroney, Phys. Rev. Lett. \textbf{110}, 120401 (2013).

\bibitem{Merm}
N.D. Mermin, Rev. Mod. Phys. \textbf{65}, 803 (1993).

\bibitem{Spekkens2005}
R. W. Spekkens, Phys. Rev. A \textbf{71}, 052108 (2005).

\bibitem{Ravi}
R. Kunjwal and R.W. Spekkens, Phys. Rev. Lett. \textbf{115}, 110403 (2015).

\bibitem{Strato}
R.L. Stratonovich, Zh. Eksp. Teor. Fiz. \textbf{31}, 1012 (1956) [Sov. Phys. JETP \textbf{4}, 891 (1957)].

\bibitem{Brif}
C. Brif, A. Mann, J. Phys. A \textbf{31},  9 (1998).

\bibitem{Wig}
E. Wigner, Phys. Rev. \textbf{40}, 749 (1932).

\bibitem{KL2}
William M. Kirby, Peter J. Love, Phys. Rev. Lett. 123, 200501 (2019).

\bibitem{HarriganSpekkens}
N.~Harrigan and R.W.~Spekkens, Found.~Phys.~\textbf{40}, 125-157 (2010).

\bibitem{Gleason}
A.M. Gleason, Indiana University Mathematics Journal \textbf{6}, 885 (1957).

\end{thebibliography}

\begin{thebibliography}{99}
	
	\bibitem{StratoSM}
	R.L. Stratonovich, Zh. Eksp. Teor. Fiz. \textbf{31}, 1012 (1956) [Sov. Phys. JETP \textbf{4}, 891 (1957)].
	
	\bibitem{BrifSM}
	C. Brif, A. Mann, J. Phys. A \textbf{31},  9 (1998).
	
	
	\bibitem{LeiferMaroney2013SM}
	M.S. Leifer and O.J.E. Maroney, Phys. Rev. Lett. \textbf{110}, 120401 (2013).
	
	\bibitem{LPSM}
	V. Chvatal, Linear programming, W. H. Freeman and Company, New York,  (1983).
	
	\bibitem{ZhuSM}
	H. Zhu, Phys. Rev. Lett. \textbf{116}, 040501 (2016).
	
	\bibitem{NegWiSM}
	V. Veitch, C. Ferrie, D. Gross, and J. Emerson, New J. Phys. \textbf{14}, 113011 (2012).
	
	\bibitem{ReWiSM} 
	N. Delfosse, P. Allard Guerin, J. Bian and R. Raussendorf, Phys. Rev. X \textbf{5}, 021003 (2015).
	
	\bibitem{Delf2SM}
	N. Delfosse, C. Okay, J. Bermejo-Vega, D. E. Browne, and R. Raussendorf, New J. Phys. \textbf{19}, 123024 (2017).
	
	\bibitem{QuWiSM}
	R. Raussendorf, D. E. Browne, N. Delfosse, C. Okay, and J. Bermejo-Vega, Phys. Rev. A \textbf{95}, 052334 (2017).
	
	\bibitem{QuWi19SM}
	R. Raussendorf, J. Bermejo-Vega, E. Tyhurst, C. Okay, M. Zurel, Phys. Rev. A \textbf{101}, 012350 (2020).
	
	\bibitem{HeimSM}
	A. Heimendahl, MSc Thesis, University of Cologne, 2019.
	
	\bibitem{KL2SM}
	William M. Kirby, Peter J. Love, Phys. Rev. Lett. 123, 200501 (2019).
	
	\bibitem{BKSM}
	S. Bravyi and A. Kitaev, Phys. Rev. A \textbf{71}, 022316 (2005).
	
	\bibitem{GomaSM}
	D. Gottesman, Proceedings of the XXII International Colloquium on Group Theoretical Methods in Physics, p. 32-43 (Cambridge, MA, International Press, 1999)
	
	\bibitem{Goma2SM}
	S. Aaronson, D. Gottesman, Phys. Rev. A \textbf{70}, 052328 (2004).
	
	\bibitem{NieChaSM}
	M.A. Nielsen and I.L. Chuang, {\em{Quantum Computation and Quantum Information}},
	Cambridge University Press (2000).
	
	\bibitem{YaoSM}
	A. Yao, Proceedings of the 34th Annual IEEE Symposium on Foundations of Computer Science, 352--361 (1993).
	
\end{thebibliography}
\end{document}